\documentclass[%
aip,
reprint,
superscriptaddress,
ymb,
prb,
twocolumn,
titlepage,
floatfix,
showpacs,
]{revtex4-2}
\usepackage{multirow} 
\usepackage{graphicx}
\usepackage{hyperref}
\usepackage{amssymb}
\usepackage{amsmath}
\usepackage{textcomp}
\usepackage{xcolor}
\allowdisplaybreaks

\definecolor{linkblue}{RGB}{49,49,148}
\hypersetup{linkcolor  = linkblue,citecolor  = linkblue,urlcolor   = linkblue,colorlinks = true,}

\makeatletter
\renewcommand*{\eqref}[1]{%
  \hyperref[{#1}]{\textup{\tagform@{\ref*{#1}}}}%
}
\makeatother

\begin{document}

\preprint{AIP/123-QED}

\title{Unveiling the nanomorphology of HfN thin films by ultrafast reciprocal space mapping}

\author{Steffen Peer Zeuschner}
\affiliation{Institut f\"ur Physik \& Astronomie,  Universit\"at Potsdam,  14476 Potsdam, Germany}

\author{Jan-Etienne Pudell}
\affiliation{European XFEL,  22869 Schenefeld,  Germany}

\author{Maximilian Mattern}
\affiliation{Institut f\"ur Physik \& Astronomie,  Universit\"at Potsdam,  14476 Potsdam, Germany}

\author{Matthias Rössle}
\affiliation{Helmholtz Zentrum Berlin, 12489 Berlin, Germany}

\author{Marc Herzog}
\affiliation{Institut f\"ur Physik \& Astronomie,  Universit\"at Potsdam,  14476 Potsdam, Germany}

\author{Andrea Baldi}
\affiliation{Department of Physics and Astronomy, Vrije Universiteit Amsterdam, 1081HV Amsterdam, Netherlands}

\author{Sven H. C. Askes}
\affiliation{Department of Physics and Astronomy, Vrije Universiteit Amsterdam, 1081HV Amsterdam, Netherlands}

\author{Matias Bargheer}
\affiliation{Institut f\"ur Physik \& Astronomie,  Universit\"at Potsdam,  14476 Potsdam, Germany}
\affiliation{Helmholtz Zentrum Berlin, 12489 Berlin, Germany}
\email{bargheer@uni-potsdam.de}

\date{\today}

\begin{abstract}
Hafnium Nitride (HfN) is a promising and very robust alternative to gold for applications of nanoscale metals. Details of the nanomorphology related to variations in strain states and optical properties can be crucial for applications in nanophotonics and plasmon-assisted chemistry. We use ultrafast reciprocal space mapping (URSM) with hard x-rays to unveil the nanomorphology of thin HfN films.
Static high-resolution x-ray diffraction reveals a twofold composition of the thin films being separated into regions with identical lattice constant and similar out-of-plane but hugely different in-plane coherence lengths.
URSM upon femtosecond laser excitation reveals different transient strain dynamics for the two respective Bragg peak components. This unambiguously locates the longer in-plane coherence length in the first 15\,nm of the thin film adjacent to the substrate. The transient shift of the broad diffraction peak displays the strain dynamics of the entire film, implying that the near-substrate region hosts nanocrystallites with small and large coherence length, whereas the upper part of the film grows in small columnar grains.
Our results illustrate that URSM is a suitable technique for non-destructive investigations of the depth-resolved nanomorphology of nanostructures. 

\end{abstract}

\newcommand{\etal}{et al.\ }
\newcommand{\ie}{i.e.,\ }
\newcommand{\eg}{e.g.\ }

\maketitle

\noindent
\section{Introduction}

Nanophotonic and plasmonic applications require materials with the highest quality.\cite{mcpe2015,vgr2020} 
Beyond their size, shape, and composition, the response of plasmonic nanostructures strongly relies on their crystalline morphology. This structure-function relationship is particularly important for lithographically-made nanostructures, in which the crystallinity of the deposited films can severely influence the nanophotonic properties \cite{mcpe2015}. In ultrafast studies the crystallinity of the system under study severely impacts the dynamics of photothermal strains.\cite{aske2019} For example, the catalysis community
developed interest in dynamically controlling strain-states in materials to modulate the binding energy of adsorbed reagents, intermediates and products, breaking the scaling relations of catalysts.\cite{arda2019}

Metal nitrides are emerging as an important class of plasmonic materials, with a range of applications such as photothermal energy conversion, plasmon-enhanced photocatalysis and nonlinear optics.\cite{diro2020,pats2018,gule2014,gule2015} HfN has proven to be very robust under large pump energy densities
which further propels metal nitride materials into the center of attention.\cite{aske2019} The ultrafast heat generation in metal nitrides has been studied before\cite{rott2023,diro2020,one2021} but not in combination with the accompanying structural dynamics from thermal expansion and ultrasound via ultrafast x-ray diffraction (UXRD). However, this information is fundamental for the understanding of the photothermal and plasmonic processes upon femtosecond laser excitation.

UXRD in general and ultrafast reciprocal space mapping (URSM) in particular are emerging analysis tools for the nanophotonics and plasmonics community to elucidate the spatio-temporal heat and strain quantitatively with a nanometer-scale precision. \cite{schi2013c,repp2016b} 
We have previously shown a variety of use cases for UXRD within the framework of picosecond ultrasonics,\cite{matt2023b} ultrafast thermometry,\cite{pude2018} nanoscale heat transport\cite{pude2020b} and plasmonic nanoparticles.\cite{repp2016b} The selectivity of UXRD for specific crystalline structures was thus far used to separately monitor the strain in individual layers within thin film heterostructures. However, more generally, URSM allows us to monitor the diffraction intensity in selected regions of reciprocal space that are specific to the crystalline morphology. In particular, we can investigate narrow and broad diffraction peaks representing the same material with different crystallite grain size. This has been exemplified for the NbO$_2$, a morphologically similar material which shows an insulator-to-metal phase transition. 
\cite{zeus2021}.

In this letter, we investigate the crystal structure of thin films of HfN 
by monitoring their ultrafast strain response upon femtosecond laser excitation using UXRD. With the combined analysis of the static and transient reciprocal space maps (RSMs), we identify two distinct morphologies within the thin films. The twofold nature of the static RSMs along the in-plane direction indicates two different crystalline coherence lengths within the HfN thin films. 
We individually probe the photo-induced out-of-plane strain dynamics of both types of crystallites to reveal the nano-morphology of the thin films. The  URSM data reveal distinct strain dynamics of the two parts of the diffraction peak. For the thickest film the narrow peak component indicates a compression whereas the broad peak component shows an expansion. This can only be rationalized by allocating the two diffraction peak features to layers with different morphology being spatially separated along the few nanometers out-of-plane direction. In addition to the URSM measurements, we show all-optical transient reflectivity measurements which indicate a sub-picosecond electron-phonon equilibration time. Based on these measurements, we model the combined photoinduced electron-phonon stress in HfN which is fed into a dynamical linear-chain model of the crystal lattice to successfully simulate the observed ultrafast thermal expansion and strain dynamics.
The two different morphologies in thin films, with the first 15 nm close to the substrate growing with higher crystalline coherence length provide an explanation for the different optical responses observed in HfN of various thicknesses. 

\noindent
\section{Results}
\subsection{Thickness dependent optical properties}
\begin{figure}[b]
\includegraphics[width = 0.5\textwidth]{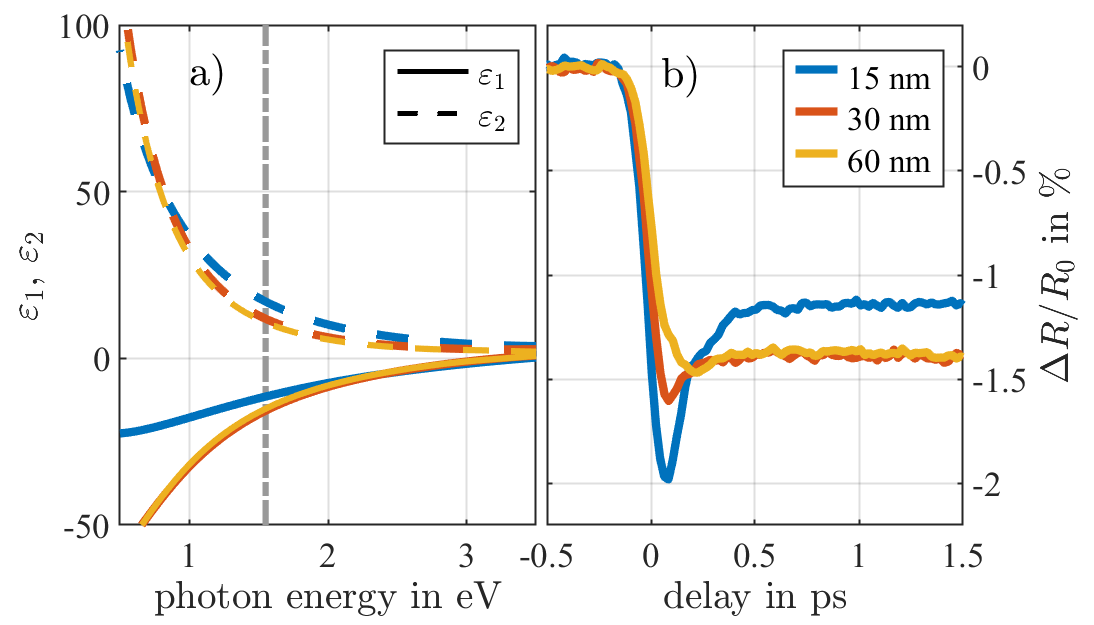}
\caption{a) Spectroscopic ellipsometry results for all three samples. Since the penetration depth is on the order of 15 nm, the retrieved dielectric constants are a property near the surface. The grey line indicates the energy 1.55 eV corresponding to the pump wavelength of 800\,nm. b) Transient relative change of the optical reflectivity averaged between 600\,nm and 650\,nm upon photoexciton with a 5\,mJ/cm$^2$ femtosecond laser pulse at 1.55\,eV. }
\label{fig:1_0}
\end{figure}

As HfN is seen as a potential candidate to replace metallic gold, e.g. for plasmonic catalysis, we briefly report how the optical properties of the HfN films depend on the thickness.
Spectroscopic ellipsometry in Fig.~\ref{fig:1_0}a) shows the real and imaginary part of the dielectric function exhibiting the characteristic negative $\varepsilon_1$ of metals. Although the data are similar for all thicknesses, the 15\,nm film clearly stands out, especially below a photon energy of 2\,eV. The dielectric function around 1.5\,eV yields a penetration depth $\delta_p=\lambda_p n/(2\pi\varepsilon_2)$ of about 15\,nm, where $\varepsilon_2$ and $n$ are the imaginary part of the dielectric function and the refractive index at the pump-wavelength $\lambda_p=800$\,nm. This implies that for thicker films the spectroscopic ellipsometry essentially measures the optical properties near the surface. In other words, the variation of optical properties across the thickness of remains inaccessible by standard spectroscopic ellipsometry.
The strong localization of plasmonically generated heat by HfN in contrast to Au suggests a very strong electron-phonon coupling \cite{aske2021}.  
Fig.~\ref{fig:1_0}b) compares results from ultrafast optical reflectivity measurements on the same 15, 30 and 60\,nm thick HfN samples pumped and probed from the front-side. The samples are excited at the pump wavelength $\lambda_p=800$\,nm with a fluence of 5\,mJ/cm$^2$. The measured reflectivity transient is probed by ultrashort white-light supercontinuum pulses, and here we show the relative cange of reflectvity around 625\,nm. The ultrafast optical response of the HfN films is very similar throughout the entire visible range from 450 to 750\,nm. Typically, such ultrafast pump-probe measurements on metals are mostly sensitive to excitations of the conduction electrons.\cite{hohl2000,wang2012b} After a fast initial drop of the reflectivity due to the strong laser-excitation of the conduction electrons, we observe a fast relaxation of the electrons with a time constant of approx. 150\,fs which we attribute to the strong electron-phonon coupling in HfN.\cite{one2021} The short spiked response attributed to the hot electrons is more pronounced in the thin film, because the diffusion into the depth of the film is reduced. We focus on short times after the excitation although our experiments reconfirm that the reflectivity drop persists for about 1\,ns.\cite{chun2017} The electrons can rapidly thermalize with high frequency optical phonons, which transport energy slowly because of their flat dispersion relation. Relaxation to acoustic phonons by standard three phonon interactions is forbidden by energy conservation.

\subsection{Sample characterization by static RSM}
In the following we report static high-resolution RSMs to establish the coexistence of a nano-columnar morphology of the HfN thin films together with a component that has a very large in-plane coherence length consistent with large crystalline slabs.

\begin{figure}[tb!]
\includegraphics[width = 0.5\textwidth]{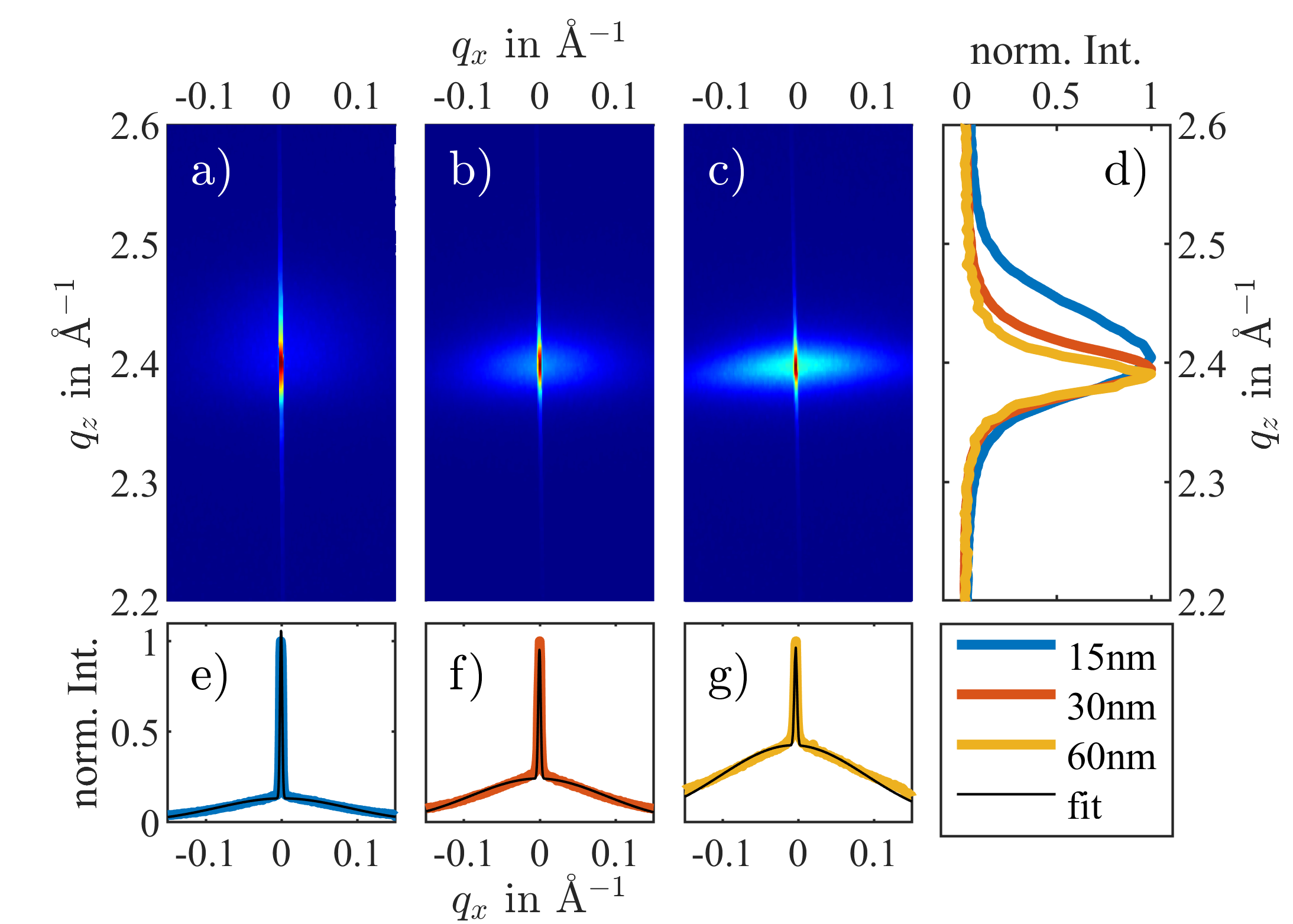}
\caption{Reciprocal space maps from XRD on the a) 15, b) 30 and c) 60\,nm thin film samples in the vicinity of the HfN(111) Bragg reflection integrated along the $q_y$ direction. d) Diffracted intensity of the broad portion of the HfN(111) Bragg reflection integrated along $q_x$. \textbf{e)-g)} Projection of the RSMs in a)-c) onto the $q_{\text{x}}$ (thick black line) axis fitted by the sum of two Gaussian functions (thin red line).}
\label{fig:I_RSM}
\end{figure}

The HfN thin films are grown on 
a c-cut sapphire substrate by reactive sputter coating (see experimental section for details). \cite{aske2019,sama1994} The three investigated samples have a HfN layer thickness of 15, 30 and 60\,nm with  preferentially
[111]-oriented rocksalt crystal structure. We investigated the HfN(111) out-of-plane Bragg reflection via static high-resolution x-ray diffraction (XRD) at the KMC3-XPP beamline at BESSY II.\cite{ross2021} The measured RSMs are depicted in Fig.~\ref{fig:I_RSM}a)-c) in which the $q_{\text{z}}$ axis corresponds to the thin films' out-of-plane direction and $q_{\text{x}}$ to their in-plane direction within the diffraction plane. Fig.~\ref{fig:I_RSM}d) shows the $q_{\text{x}}$-integrated intensity distributions along $q_{\text{z}}$.
The position of the HfN(111) Bragg reflection at $q_{\text{z}} \approx 2.4$\,\AA$^{\textrm{-}1}$ yields a cubic lattice constant of $a_0 \approx 4.53$\,\AA{}, which is consistent with the bulk lattice constant of HfN.\cite{seo2004} Clearly, the reflection exhibits two contributions of a broad and narrow peak along $q_{\text{x}}$ as illustrated by the $q_{\text{z}}$-integrated intensity profiles depicted in Fig.~\ref{fig:I_RSM}e)-g) . The two portions are fitted separately with Gaussian functions along the $q_{\text{x}}$ and $q_{\text{z}}$-direction to yield the widths $\sigma_{\text{x,b}}$ and $\sigma_{\text{z,b}}$ as well as $\sigma_{\text{x,n}}$ and $\sigma_{\text{z,n}}$ for the broad and narrow peak, respectively. From this, the corresponding coherence lengths are calculated via $L=\pi/\sqrt{2ln2}\sigma$, which are summarized in Tab.~\ref{tab:RSM}. The widths of the narrow portion of the peak translate to a coherence length of 170\,nm in-plane and 15\,nm out-of-plane for all three samples. The broad portion of the peak gets broader along the $q_{\text{z}}$ direction with decreasing layer thickness, visualized by the $q_{\text{z}}$ projections of the RSMs in Fig.\ref{fig:I_RSM} d). This relates to a component with small coherence length of 3\,nm in-plane for all samples and an out of plane coherence length of 7, 12, and 15 nm for the 15, 30, and 60 nm films, respectively. 
\begin{table}[b]
\renewcommand{\arraystretch}{2}
\begin{tabular}{|c|cccc|cccc|}
\hline
 & \multicolumn{4}{c|}{broad peak portion}                                                                                      & \multicolumn{4}{c|}{narrow peak portion}                                                                                                                        \\ \cline{2-9} 
 & \multicolumn{1}{c|}{\large $\frac{\sigma_{\text{z}}}{\text{\footnotesize\AA$^{\textrm{-}1}$}}$ } & \multicolumn{1}{c|}{\large$\frac{L_{\perp}}{\text{ \footnotesize nm}}$} & \multicolumn{1}{c|}{\large$\frac{\sigma_{\text{x}}}{\text{\footnotesize\AA$^{\textrm{-}1}$}}$}                  &\large$\frac{L_{\parallel}}{\text{ \footnotesize nm}}$          & \multicolumn{1}{c|}{\large$\frac{\sigma_{\text{z}}}{\text{\footnotesize\AA$^{\textrm{-}1}$}}$}                  & \multicolumn{1}{c|}{\large$\frac{L_{\perp}}{\text{\footnotesize nm}}$}                  & \multicolumn{1}{c|}{\large$\frac{\sigma_{\text{x}}}{\text{\footnotesize\AA$^{\textrm{-}1}$}}$}       &\large$\frac{L_{\parallel}}{\text{\footnotesize nm}}$       \\ \hline
 15\,nm film& \multicolumn{1}{c|}{0.038} & \multicolumn{1}{c|}{7} & \multicolumn{1}{c|}{\multirow{3}{*}{$0.09$}} & \multirow{3}{*}{3} & \multicolumn{1}{c|}{0.017} & \multicolumn{1}{c|}{16} & \multicolumn{1}{c|}{\multirow{3}{*}{0.0016}} & \multirow{3}{*}{170} \\ \cline{1-3} \cline{6-7}
 30\,nm film& \multicolumn{1}{c|}{0.022} & \multicolumn{1}{c|}{12} & \multicolumn{1}{c|}{}                  &                   & \multicolumn{1}{c|}{0.017}                  & \multicolumn{1}{c|}{16}                  & \multicolumn{1}{l|}{}                  &                   \\ \cline{1-3} \cline{6-7}
 60\,nm film& \multicolumn{1}{c|}{0.018} & \multicolumn{1}{c|}{15} & \multicolumn{1}{c|}{}                  &                   & \multicolumn{1}{c|}{0.014}                   & \multicolumn{1}{c|}{19}                  & \multicolumn{1}{l|}{}                  &         \\ \hline         
\end{tabular}
\caption{Widths $\sigma_{z/x}$ of the measured broad and narrow portions of the HfN(111) diffraction peaks along the $q_{z/x}$ direction for the 15, 30 and 60\,nm thin films extracted by Gaussian fitting. The coherence lengths $L_{\perp}$ (out-of-plane) and $L_{\parallel}$ (in-plane) correspond to the $\sigma_{z/x}$ in the neighbouring column.}
\label{tab:RSM}
\end{table}

The RSMs thus indicate two 
different species or grains of HfN in the thin films. Since the $q_{\text{z}}$ position is identical, both components must have a [111]-orientation. The coherence lengths show that the portion of the film which contributes the narrow diffraction peak has spatial dimensions of 15\,nm out-of-plane and 170\,nm in-plane suggesting slab-like shape with relatively high crystalline quality. The coherence lengths of the broad peak are comparably small. Moreover, the increasing diffraction intensity of the broad component is consistent with the increase of the scattering volume with rising layer thickness. Thus, the broad portion of the diffraction peak presumably originates from a columnar morphology of the thin film, which has been reported before.\cite{pats2018} Also, the coherence length corresponding to the narrow peak does not change with increasing layer thickness. This already suggests assigning this unchanged fraction to be the 15\,nm of each film, which are adjacent to the substrate.  \\

\noindent
\subsection{UXRD analysis of strain dynamics}

\begin{figure}[tb!]
\includegraphics[width = 0.5\textwidth]{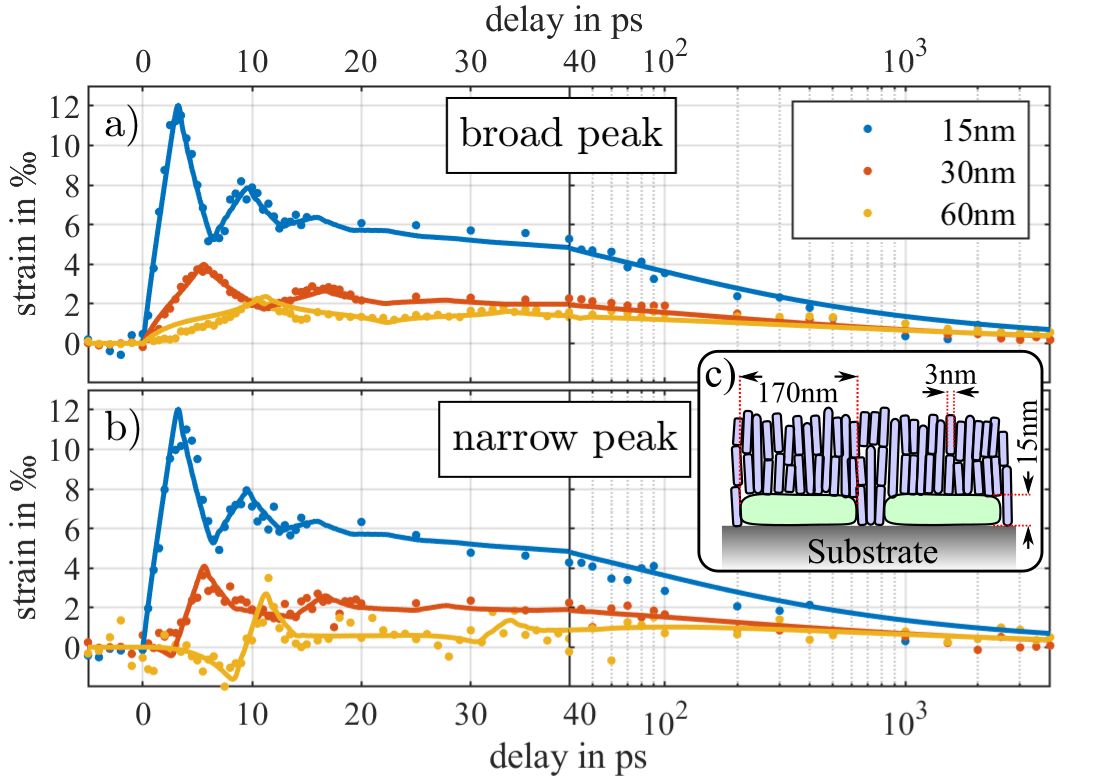}
\caption{Time-resolved average strain in thin HfN films measured via URSM with selective monitoring of the a) broad and b) narrow portion of the HfN(111) Bragg peak. The solid lines represent the corresponding modeled strain calculated with the udkm1Dsim toolbox averaged over a) the entire HfN layer thickness and b) the bottom, near-substrate 15\,nm of the respective HfN layer. c) Pictorial sketch of a vertical cross section of the thin film illustrating the mixed nanomorphology of slab-like and columnar crystallites in the HfN films as deduced from the static and ultrafast XRD studies.}
\label{fig:II_TRstrain}
\end{figure}

The structural insights derived from static RSM measurements are confirmed by an intriguing observation in the UXRD measurements which probe the transient strain dynamics. In the following we demonstrate that the two different crystalline morphologies show different ultrafast strain dynamics, because the component with large coherence length - probably mediated by the substrate - is exclusively located in the first few nanometers of the subtrate.
In the UXRD experiment, the thin films are excited by pump pulses with a central wavelength of 800\,nm, a duration of approximately 100\,fs and a pump fluence of 8.5\,mJ/cm$^2$.
RSMs similar to those presented in Fig.~\ref{fig:I_RSM} are measured as a function of the pump-probe delay $t$. The $q_{\text{z}}(t)$ positions of the broad and narrow Bragg peak components are separately extracted according to their different $q_{\text{x}}$ (see experimental section for details). 
For both morphological components, the average out-of-plane strain $\eta_{\text{z}}(t)$ is calculated from the relative change of the out-of-plane lattice constant $a$, which is given by Bragg's law:
\begin{equation}
\label{eq:1}
    \eta_{\text{z}}(t) =\frac{\Delta a(t)}{a_0} = \frac{q_{\text{z}}(t_0)-q_{\text{z}}(t)}{q_{\text{z}}(t_0)}.
\end{equation}
The strain transients of the broad and narrow peak for all three samples are depicted in Fig.~\ref{fig:II_TRstrain}a) and b), respectively. We find that the narrow and broad diffraction peak components exhibit significantly different, even opposing motions (yellow lines in Fig.~\ref{fig:II_TRstrain}a) and b)) in reciprocal space upon femtosecond laser excitation. 

We first discuss the average strain derived from the broad component in Fig.~\ref{fig:II_TRstrain}a), which exhibits the conventional ultrafast response of laser-excited thin film transducers that converts light into heat and strain.\cite{thom1986,schi2014b,matt2023b} The absorption of the pump pulse rapidly increases the energy density inside the HfN thin film. The proportional rise of the thermoelastic stress is relaxed by a rapid expansion (strain) of the laser-heated part of the film. Averaging over initial oscillations, the average strain in the thin film in the quasi-static strain limit is given by the thermal expansion of the film under the epitaxial constraint of clamped in-plane lattice constants.\cite{matt2023b} On nanosecond timescales heat diffuses into the adjacent substrate, resulting in a relaxation of the out-of-plane strain. The damped oscillations of the average strain during the first 20\,ps is a result of the acoustic wavepacket that is triggered by the quasi-instantaneous laser-induced stress rise and traverses the thin film at the longitudinal speed of sound $v_{\text{s}}=5.4$\,nm/ps \cite{demira2018}. This wavepacket is partially reflected at the HfN-Sapphire-interface due to an acoustic impedance mismatch. It thus bounces back and forth within the layer a few times, resulting in the strongly damped oscillation. The timings $T_{\text{max}}=d/v_{\text{s}}\approx3$\,ps, 6\,ps and 12\,ps of the first maxima of the strain transients is defined by the corresponding layer thicknesses $d=15$\,nm, 30\,nm and 60\,nm as well as the sound velocity $v_{\text{s}}$. The sharp interface results in sharp strain fronts that give rise to the characteristic triangular form of the measured average strain. 
The optical constants of HfN at the pump wavelength of 800\,nm imply an optical penetration depth of approx. $\delta_p= 15$\,nm. That is, only 15-20\,nm of sub-surface HfN are directly excited by the pump pulse. This results in a strongly inhomogeneous excitation profile for the thicker films ($d=30$, 60\,nm) where the substrate-near part of the film thus remains unexcited. Hence, the generated strain pulse is launched at the surface and has a rather small extension compared to the film thickness. For the thinnest film ($d=15$\,nm), however, a rather homogeneous excitation profile is calculated.\cite{schi2021}

We now focus on the strain derived from the narrow portion of the Bragg reflection (Fig.~\ref{fig:II_TRstrain}b) and compare it to its broad counterpart (Fig.~\ref{fig:II_TRstrain}a).
At late time delays beyond 100\,ps, both components exhibit identical quasi-static responses of a thermally expanded lattice which relaxes on nanosecond timescales due to heat diffusion into the substrate.
In case of the 15\,nm film, both components reveal also identical behaviour of the transient strain at early time delays. However, the narrow-peak response of the 30\,nm film seems delayed with respect to the broad-peak response and in the 60\,nm film the narrow peak even shows a negative strain response within the first 10\,ps (Fig.~\ref{fig:II_TRstrain}b) corresponding to a compression of the laser-excited film along the out-of-plane direction. 
Such compression effects have been previously observed in multilayers
where the impulsive expansion of a laser-excited top metal layer leads to an initial compression of a buried detection layer.\cite{pude2018,zeus2019,repp2020a,deb2021b,matt2023b} The concept of compressing crystalline layers by laser-excitation of adjacent transducers was even exploited in ultrafast x-ray optics to as a switchable Bragg mirror.\cite{gaal2012,gaal2014,sand2016}
Our findings thus readily suggest that the portion of the HfN thin film being responsible for the narrow Bragg peak, i.e. with long in-plane coherence length, is in fact a nanosize fraction of the film adjacent to the substrate. The marked difference to previous experiments is that the compressed layer is not an individual detection layer, but it is made from the same material with the same out-of-plane lattice constant. It can only be distinguished from the remaining part of the HfN thin film due to the different in-plane coherence length, which gives rise to diffraction intensity in a separate part of reciprocal space.

We note here that the mere qualitative analysis of the UXRD signals readily implies an approximate morphology of the HfN thin film as sketched in Fig.~\ref{fig:II_TRstrain}c). On the one hand, HfN slabs with a large in-plane coherence length are located at the substrate interface. On the other hand, a columnar nanomorphology with very short in-plane coherence length extends throughout the thickness of the film. The latter is evidenced by the timing of the strain maxima in Fig.~\ref{fig:II_TRstrain}a, which implies that the broad-peak response originates from the acoustic dynamics across the entire thickness of the film.


\section{Modeling}
\begin{figure}[tb!]
\includegraphics[width = 0.45\textwidth]{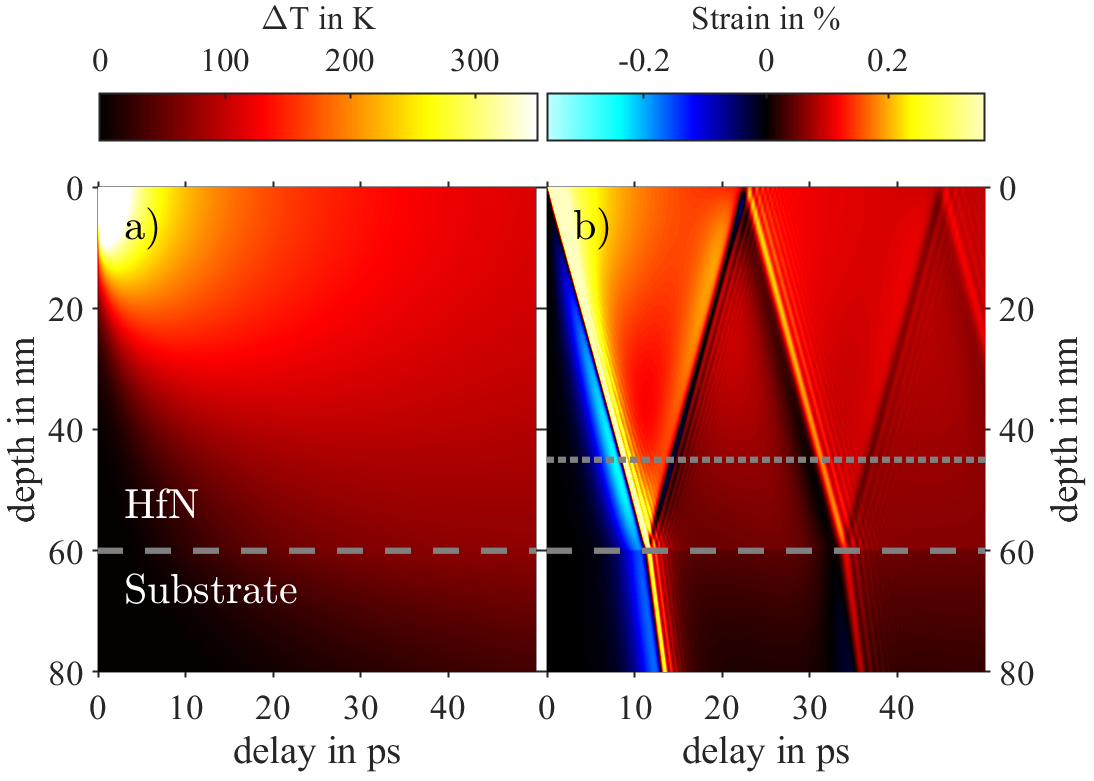}
\caption{a) Modeled spatio-temporal temperature map of the 60\,nm HfN thin film upon photoexciton with a 8.5\,mJ/cm$^2$ femtosecond laser pulse at 800\,nm wavelength. b) Spatio-temporal strain map resulting from the temperature map shown in panel a). The long- and short-dashed lines in a) and b) indicate the HfN-Substrate interface and the area of the bottom 15\,nm of the thin film, respectively.}
\label{fig:III_TRstrain}
\end{figure}

To support this interpretation and give a quantitative perspective, we model the data utilizing the library \textsc{udkm1Dsim}, a simulation toolkit for ultrafast electron and phonon transport as well as the corresponding temperatures and strain dynamics in layered materials.\cite{schi2014,schi2021} The results of the modeling shown as solid lines in Fig.~\ref{fig:II_TRstrain} fit the experimental data very well. A proper description of electron-phonon dynamics in laser-excited metals often requires two-temperature models.\cite{hohl2000,wang2012b,pude2018,herz2022} 
The electron-phonon relaxation in HfN occurs on a 100-fs timescale (see Fig.~\ref{fig:1_0}b)), which is much faster than any other process that our experiment is sensitive to. It is in agreement with previous ultrafast transient absorption spectroscopy studies on water-dispersed HfN nanoparticles.\cite{one2021} This suggests that the use of a simpler one-temperature model suffices. 
The optical as well as other thermophysical properties of HfN and the sapphire substrate listed in Tab. \ref{tab:props} are used in the udkm1Dsim toolbox to calculate the spatio-temporal temperature distribution that is initiated by the absorbed laser pulse energy and evolves according to the laws of heat diffusion (Fig.~\ref{fig:III_TRstrain}a)). The thermal energy generates a transient local stress that triggers spatio-temporal strain fields, which are modeled by the toolbox utilizing a linear-chain model. Figure~\ref{fig:III_TRstrain}b) exemplifies the resulting strain map for the 60\,nm thick HfN film.

\begin{table}
\renewcommand{\arraystretch}{2}
    \centering
    \begin{tabular}{|l|c|c|}
    \hline
         & HfN  & Al$_2$O$_3$\\
         \hline
heat capacity (J/kg/K)& 207 \cite{saha2010} & 778 \cite{ginn1953}\\
         \hline
thermal conductivity (W/m/K)& 90 \cite{li2020} & 40 \cite{dobr2009} \\
         \hline
linear thermal expansion ($10^{-6}$/K)& 8 \cite{Aigner.1994} & 6.6 \cite{luch2003} \\
         \hline
sound velocity (nm/ps)& 5.4 \cite{demira2018} & 11.1 \cite{wach1960}\\
         \hline
    \end{tabular}
    \caption{Thermophysical parameters used as input for the one-temperature model and linear chain model of the udkm1Dsim-toolbox to calculate the spatio-temporal temperature and strain fields in the HfN samples upon femtosecond laser-excitation, see Fig. \ref{fig:III_TRstrain}}
    \label{tab:props}
\end{table}
The temperature and strain maps (Fig.~\ref{fig:III_TRstrain}a)-b)) clearly illustrate that the thermoelastic stress inscribed by the optical excitation is strongest at the HfN surface resulting in the aforementioned expansion of this surface-near part of the thin film (red and yellow shades) and a concomitant compression of the deeper-lying parts until approx. 10\,ps (blue shades).
After the stress release in the top unit cells the expansion wave proceeds deeper into the thin film forming a bipolar strain pulse with a compressive leading part and a tensile trailing part.\cite{thom1986,schi2014b,matt2023b} In particular, the strain map shows the dynamics of the bipolar strain pulse bouncing back and forth in the thin film.

The calculated strain maps can now be used for comparison with the UXRD data as the latter is a direct measurement of the average transient strain in the laser-excited HfN thin films. Spatially integrating the strain distribution over the whole respective layer thickness for each time delay yields the modeled average strain for each of the thin films. The corresponding results are plotted as solid lines in Fig.~\ref{fig:II_TRstrain}a). The model reproduces all three data sets obtained from the broad Bragg peaks with high accuracy.

According to the implications from the above qualitative discussion of the UXRD data, we indeed find that we are able to fit the narrow-peak strain transients by averaging the strain maps over the bottom 15\,nm of the HfN thin films. This region is indicated by two horizontal lines in Fig.~\ref{fig:III_TRstrain}b). The results of this particular procedure are depicted as solid lines in Fig.~\ref{fig:II_TRstrain}b). For the thinnest film, the strain transients in Fig.~\ref{fig:II_TRstrain}a) and b) are practically identical as the bottom 15\,nm represent the entire film already. In the 60\,nm film, the bottom part is initially compressed due to the leading compressive part of the bipolar strain pulse launched at the surface, followed by the tensile trailing part at about 10\,ps. The 30\,nm film represents an intermediate case where the increasing average strain in the bottom part is only compensated by the initial compression of the bipolar strain pulse within the first few picoseconds.

We note here that while the strain transients of the 30 and 60\,nm films presented in Fig.\ref{fig:II_TRstrain}a) and b) are fully reproduced by the modeling using the experimental pump fluence of 8.5\,mJ/cm$^2$,  the 15\,nm film transients required an upscaling by a factor of 1.6. The reason for this significantly enhanced expansion in the 15\,nm sample is still under investigation. Possible reasons could be related to the peculiar morphology identified in this study potentially resulting in enhanced local absorption or increased out-of-plane strain response due to the epitaxial constraints from the substrate (Poisson effect).\cite{repp2018,matt2023b}

\noindent
\section{Conclusions}

The presented experimental results and modeling lead to the following interpretation. The investigated thin HfN films consist of two distinct morphologies or crystallites, evidenced by the twofold shape of the measured Bragg peaks along the in-plane reciprocal direction. The narrow peak component originates from the bottom 15\,nm near the substrate for all HfN films. This is evidenced by both the essentially constant diffraction intensity of this Bragg peak component for all samples and the modeling of its strain dynamics revealed by UXRD. 
The modeling of the transient strain response of the broad Bragg peak component implies that it must originate from regions of the HfN film spread across the entire film thickness. This is also consistent with the static XRD intensity of this component being proportional to the film thickness.
We thus propose a nanoscale morphology of the thin HfN films as sketched in Fig.~\ref{fig:II_TRstrain}c). In close proximity to the substrate the HfN films host slab-like crystallites (green) with a lateral dimension of a few hundred nanometers and a height of about 15\,nm. In between and on top of those nanoslabs, the HfN films consist of columnar crystallites with only a few nanometer width which has been observed to be typical for transition metal nitrides.\cite{pats2018}

On the technological side, we have demonstrated that UXRD, in particular URSM, is a highly capable and useful tool to complement conventional XRD in structural analysis of nanoscale heterostructures even within a thin film of a single material. The depth sensitivity of the URSM experiments is encoded in the temporal signatures of the transient strain following laser pulse excitation which is extracted from the transient RSM data. 
Finally, our investigations shed light on the temperature and strain dynamics of HfN nanostructures upon femtosecond laser pulse excitation which is vital for pursuing to understand and interpret future investigations on nitride-based nanophotonics, nanoscale thermal properties and plasmon-assisted catalysis.

\section{Experimental section}
\subsection{Deposition of HfN on sapphire substrates}
The sapphire substrates (c-cut, [0001]-oriented) were first cleaned for 20\,min using a base piranha solution (5:1:1 solution of distilled H$_2$O, 30\% ammonia, and 30\% H$_2$O$_2$) at 60$^\circ$C, then rinsed extensively using distilled H$_2$O and isopropanol, and blow-dried using pressurized N$_2$. HfN was deposited using a Flextura reactive sputter coating system (Polyteknik AS, Denmark) with a base pressure of $\leq 9\cdot10^{-8}$\,mbar. A pure Hf target was used, which was sputter cleaned for at least 2 minutes directly prior to deposition (i.e. during target cleaning the shutter in front of the sample was closed). The substrate was rotated at 4 rpm during deposition and neither heated nor cooled. HfN was deposited at a typical rate of 8.5 nm/min (determined using a profilometer) using radio frequency (RF) magnetron sputtering at 150\,W power at a pressure of $2.6 \cdot 10^{-3}$\,mbar, in a gas mixture of 35.3\,sccm Ar and 0.7\,sccm N$_2$. The produced HfN coatings are optically reflective with a light-yellow metallic appearance and are progressively more opaque with increasing layer thickness

\subsection{Transient reflectivity}
The optical pump-probe setup utilized 150\,fs pump pulses at 800\,nm from a Spectra-Physics Spitfire regenartive amplifier. The white-light supercontinuum probe pulses were generated by focusing weak 800\,nm pulses into a sapphire plate and were subsequently focused onto the samples down to a beam diameter of about 50 µm (FWHM), more than ten times smaller than the size of the pump beam to ensure a sufficiently homogeneous excitation across the probed volume. 

\subsection{UXRD}
For the laser-based UXRD experiments we use our laser-driven plasma x-ray source at the University of Potsdam, which generates Cu-K$_{\alpha}$ hard x-ray pulses with a few hundred femtoseconds pulse duration.\cite{schi2012} The x-ray beam size on the sample was approximately 250\,x\,250\,µm$^2$ (FWHM), four times smaller than the size of the pump beam to ensure a sufficiently homogeneous excitation across the probed volume. The RSMs are measured in a symmetric diffraction geometry by scanning the diffraction angles $\omega$ and $2\theta$ and transforming the 2D images of a Pilatus 100k area detector (Dectris) into $q$-space.\cite{schi2012,schi2013c} The same procedure is used at the KMC3-XPP beamline at BESSY II for the static RSMs.\cite{ross2021}

We monitor the $q_{\text{z}}$-positions of the two components of the Bragg peak as a function of the pump-probe delay $t$ by integrating the RSMs across separated areas along $q_{\text{x}}$ and fitting the resulting intensity distributions with Gaussian functions. To discriminate the two components, the integration boundaries are set to $-0.004$ \AA$^{\textrm{-}1} < q_{\text{x}} < 0.004$ \AA$^{\textrm{-}1}$ for the narrow portion and the rest is used for the broad portion.

\noindent
\textbf{Acknowledgements}

This research is driven by the Deutsche Forschungsgemeinschaft (DFG, German Research Foundation) – CRC/SFB 1636 – Project ID 510943930 - Projects No. A01, Z02 and the Mercator fellowship of A.B.. A.B. gratefully acknowledges the Dutch Research Council (NWO) for the Vidi 680-47-550 grant.

\bibliographystyle{achemso}

\begin{mcitethebibliography}{44}
\providecommand*\natexlab[1]{#1}
\providecommand*\mciteSetBstSublistMode[1]{}
\providecommand*\mciteSetBstMaxWidthForm[2]{}
\providecommand*\mciteBstWouldAddEndPuncttrue
  {\def\EndOfBibitem{\unskip.}}
\providecommand*\mciteBstWouldAddEndPunctfalse
  {\let\EndOfBibitem\relax}
\providecommand*\mciteSetBstMidEndSepPunct[3]{}
\providecommand*\mciteSetBstSublistLabelBeginEnd[3]{}
\providecommand*\EndOfBibitem{}
\mciteSetBstSublistMode{f}
\mciteSetBstMaxWidthForm{subitem}{(\alph{mcitesubitemcount})}
\mciteSetBstSublistLabelBeginEnd
  {\mcitemaxwidthsubitemform\space}
  {\relax}
  {\relax}

\bibitem[McPeak \latin{et~al.}(2015)McPeak, Jayanti, Kress, Meyer, Iotti, Rossinelli, and Norris]{mcpe2015}
McPeak,~K.~M.; Jayanti,~S.~V.; Kress,~S. J.~P.; Meyer,~S.; Iotti,~S.; Rossinelli,~A.; Norris,~D.~J. Plasmonic Films Can Easily Be Better: Rules and Recipes. \emph{ACS Photonics} \textbf{2015}, \emph{2}, 326--333\relax
\mciteBstWouldAddEndPuncttrue
\mciteSetBstMidEndSepPunct{\mcitedefaultmidpunct}
{\mcitedefaultendpunct}{\mcitedefaultseppunct}\relax
\EndOfBibitem
\bibitem[{V Grayli} \latin{et~al.}(2020){V Grayli}, Zhang, MacNab, Kamal, Star, and Leach]{vgr2020}
{V Grayli},~S.; Zhang,~X.; MacNab,~F.~C.; Kamal,~S.; Star,~D.; Leach,~G.~W. Scalable, Green Fabrication of Single-Crystal Noble Metal Films and Nanostructures for Low-Loss Nanotechnology Applications. \emph{ACS nano} \textbf{2020}, \emph{14}, 7581--7592\relax
\mciteBstWouldAddEndPuncttrue
\mciteSetBstMidEndSepPunct{\mcitedefaultmidpunct}
{\mcitedefaultendpunct}{\mcitedefaultseppunct}\relax
\EndOfBibitem
\bibitem[Askes \latin{et~al.}(2019)Askes, Schilder, Zoethout, Polman, and Garnett]{aske2019}
Askes,~S. H.~C.; Schilder,~N.~J.; Zoethout,~E.; Polman,~A.; Garnett,~E.~C. Tunable plasmonic HfN nanoparticles and arrays. \emph{Nanoscale} \textbf{2019}, \emph{11}, 20252--20260\relax
\mciteBstWouldAddEndPuncttrue
\mciteSetBstMidEndSepPunct{\mcitedefaultmidpunct}
{\mcitedefaultendpunct}{\mcitedefaultseppunct}\relax
\EndOfBibitem
\bibitem[Ardagh \latin{et~al.}(2019)Ardagh, Abdelrahman, and Dauenhauer]{arda2019}
Ardagh,~M.~A.; Abdelrahman,~O.~A.; Dauenhauer,~P.~J. Principles of Dynamic Heterogeneous Catalysis: Surface Resonance and Turnover Frequency Response. \emph{ACS Catalysis} \textbf{2019}, \emph{9}, 6929--6937\relax
\mciteBstWouldAddEndPuncttrue
\mciteSetBstMidEndSepPunct{\mcitedefaultmidpunct}
{\mcitedefaultendpunct}{\mcitedefaultseppunct}\relax
\EndOfBibitem
\bibitem[Diroll \latin{et~al.}(2020)Diroll, Saha, Shalaev, Boltasseva, and Schaller]{diro2020}
Diroll,~B.~T.; Saha,~S.; Shalaev,~V.~M.; Boltasseva,~A.; Schaller,~R.~D. Broadband Ultrafast Dynamics of Refractory Metals: TiN and ZrN. \emph{Advanced Optical Materials} \textbf{2020}, \emph{8}, 2000652\relax
\mciteBstWouldAddEndPuncttrue
\mciteSetBstMidEndSepPunct{\mcitedefaultmidpunct}
{\mcitedefaultendpunct}{\mcitedefaultseppunct}\relax
\EndOfBibitem
\bibitem[Patsalas \latin{et~al.}(2018)Patsalas, Kalfagiannis, Kassavetis, Abadias, Bellas, Lekka, and Lidorikis]{pats2018}
Patsalas,~P.; Kalfagiannis,~N.; Kassavetis,~S.; Abadias,~G.; Bellas,~D.~V.; Lekka,~C.; Lidorikis,~E. Conductive nitrides: Growth principles, optical and electronic properties, and their perspectives in photonics and plasmonics. \emph{Materials Science and Engineering: R: Reports} \textbf{2018}, \emph{123}, 1--55\relax
\mciteBstWouldAddEndPuncttrue
\mciteSetBstMidEndSepPunct{\mcitedefaultmidpunct}
{\mcitedefaultendpunct}{\mcitedefaultseppunct}\relax
\EndOfBibitem
\bibitem[Guler \latin{et~al.}(2014)Guler, Boltasseva, and Shalaev]{gule2014}
Guler,~U.; Boltasseva,~A.; Shalaev,~V.~M. Applied physics. Refractory plasmonics. \emph{Science} \textbf{2014}, \emph{344}, 263--264\relax
\mciteBstWouldAddEndPuncttrue
\mciteSetBstMidEndSepPunct{\mcitedefaultmidpunct}
{\mcitedefaultendpunct}{\mcitedefaultseppunct}\relax
\EndOfBibitem
\bibitem[Guler \latin{et~al.}(2015)Guler, Shalaev, and Boltasseva]{gule2015}
Guler,~U.; Shalaev,~V.~M.; Boltasseva,~A. Nanoparticle plasmonics: going practical with transition metal nitrides. \emph{Mater. Today} \textbf{2015}, \emph{18}, 227--237\relax
\mciteBstWouldAddEndPuncttrue
\mciteSetBstMidEndSepPunct{\mcitedefaultmidpunct}
{\mcitedefaultendpunct}{\mcitedefaultseppunct}\relax
\EndOfBibitem
\bibitem[{Rotta Loria} \latin{et~al.}(2023){Rotta Loria}, Bricchi, Schirato, Mascaretti, Mancarella, Naldoni, {Li Bassi}, {Della Valle}, and Zavelani-Rossi]{rott2023}
{Rotta Loria},~S.; Bricchi,~B.~R.; Schirato,~A.; Mascaretti,~L.; Mancarella,~C.; Naldoni,~A.; {Li Bassi},~A.; {Della Valle},~G.; Zavelani-Rossi,~M. Unfolding the Origin of the Ultrafast Optical Response of Titanium Nitride. \emph{Advanced Optical Materials} \textbf{2023}, \relax
\mciteBstWouldAddEndPunctfalse
\mciteSetBstMidEndSepPunct{\mcitedefaultmidpunct}
{}{\mcitedefaultseppunct}\relax
\EndOfBibitem
\bibitem[O'Neill \latin{et~al.}(2021)O'Neill, Frehan, Zhu, Zoethout, Mul, Garnett, Huijser, and Askes]{one2021}
O'Neill,~D.~B.; Frehan,~S.~K.; Zhu,~K.; Zoethout,~E.; Mul,~G.; Garnett,~E.~C.; Huijser,~A.; Askes,~S. H.~C. Ultrafast Photoinduced Heat Generation by Plasmonic HfN Nanoparticles. \emph{Advanced Optical Materials} \textbf{2021}, \emph{9}, 2100510\relax
\mciteBstWouldAddEndPuncttrue
\mciteSetBstMidEndSepPunct{\mcitedefaultmidpunct}
{\mcitedefaultendpunct}{\mcitedefaultseppunct}\relax
\EndOfBibitem
\bibitem[Schick \latin{et~al.}(2013)Schick, Shayduk, Bojahr, Herzog, von {Korff Schmising}, Gaal, and Bargheer]{schi2013c}
Schick,~D.; Shayduk,~R.; Bojahr,~A.; Herzog,~M.; von {Korff Schmising},~C.; Gaal,~P.; Bargheer,~M. Ultrafast reciprocal-space mapping with a convergent beam. \emph{Journal of Applied Crystallography} \textbf{2013}, \emph{46}, 1372--1377\relax
\mciteBstWouldAddEndPuncttrue
\mciteSetBstMidEndSepPunct{\mcitedefaultmidpunct}
{\mcitedefaultendpunct}{\mcitedefaultseppunct}\relax
\EndOfBibitem
\bibitem[von Reppert \latin{et~al.}(2016)von Reppert, Sarhan, Stete, Pudell, {Del Fatti}, Crut, Koetz, Liebig, Prietzel, and Bargheer]{repp2016b}
von Reppert,~A.; Sarhan,~R.~M.; Stete,~F.; Pudell,~J.; {Del Fatti},~N.; Crut,~A.; Koetz,~J.; Liebig,~F.; Prietzel,~C.; Bargheer,~M. Watching the Vibration and Cooling of Ultrathin Gold Nanotriangles by Ultrafast X-ray Diffraction. \emph{The Journal of Physical Chemistry C} \textbf{2016}, \emph{120}, 28894--28899\relax
\mciteBstWouldAddEndPuncttrue
\mciteSetBstMidEndSepPunct{\mcitedefaultmidpunct}
{\mcitedefaultendpunct}{\mcitedefaultseppunct}\relax
\EndOfBibitem
\bibitem[Mattern \latin{et~al.}(2023)Mattern, von Reppert, Zeuschner, Herzog, Pudell, and Bargheer]{matt2023b}
Mattern,~M.; von Reppert,~A.; Zeuschner,~S.~P.; Herzog,~M.; Pudell,~J.-E.; Bargheer,~M. Concepts and use cases for picosecond ultrasonics with x-rays. \emph{Photoacoustics} \textbf{2023}, \emph{31}, 100503\relax
\mciteBstWouldAddEndPuncttrue
\mciteSetBstMidEndSepPunct{\mcitedefaultmidpunct}
{\mcitedefaultendpunct}{\mcitedefaultseppunct}\relax
\EndOfBibitem
\bibitem[Pudell \latin{et~al.}(2018)Pudell, Maznev, Herzog, Kronseder, Back, Malinowski, von Reppert, and Bargheer]{pude2018}
Pudell,~J.; Maznev,~A.~A.; Herzog,~M.; Kronseder,~M.; Back,~C.~H.; Malinowski,~G.; von Reppert,~A.; Bargheer,~M. Layer specific observation of slow thermal equilibration in ultrathin metallic nanostructures by femtosecond X-ray diffraction. \emph{Nature Communications} \textbf{2018}, \emph{9}, 3335\relax
\mciteBstWouldAddEndPuncttrue
\mciteSetBstMidEndSepPunct{\mcitedefaultmidpunct}
{\mcitedefaultendpunct}{\mcitedefaultseppunct}\relax
\EndOfBibitem
\bibitem[Pudell \latin{et~al.}(2020)Pudell, Mattern, Hehn, Malinowski, Herzog, and Bargheer]{pude2020b}
Pudell,~J.-E.; Mattern,~M.; Hehn,~M.; Malinowski,~G.; Herzog,~M.; Bargheer,~M. Heat Transport without Heating?---An Ultrafast X--Ray Perspective into a Metal Heterostructure. \emph{Advanced Functional Materials} \textbf{2020}, \emph{30}, 2004555\relax
\mciteBstWouldAddEndPuncttrue
\mciteSetBstMidEndSepPunct{\mcitedefaultmidpunct}
{\mcitedefaultendpunct}{\mcitedefaultseppunct}\relax
\EndOfBibitem
\bibitem[Zeuschner \latin{et~al.}(2021)Zeuschner, Mattern, Pudell, von Reppert, R{\"o}ssle, Leitenberger, Schwarzkopf, Boschker, Herzog, and Bargheer]{zeus2021}
Zeuschner,~S.~P.; Mattern,~M.; Pudell,~J.-E.; von Reppert,~A.; R{\"o}ssle,~M.; Leitenberger,~W.; Schwarzkopf,~J.; Boschker,~J.~E.; Herzog,~M.; Bargheer,~M. Reciprocal space slicing: A time-efficient approach to femtosecond x-ray diffraction. \emph{Structural dynamics (Melville, N.Y.)} \textbf{2021}, \emph{8}, 014302\relax
\mciteBstWouldAddEndPuncttrue
\mciteSetBstMidEndSepPunct{\mcitedefaultmidpunct}
{\mcitedefaultendpunct}{\mcitedefaultseppunct}\relax
\EndOfBibitem
\bibitem[Askes and Garnett(2021)Askes, and Garnett]{aske2021}
Askes,~S.~H.; Garnett,~E.~C. Ultrafast thermal imprinting of plasmonic hotspots. \emph{Advanced Materials} \textbf{2021}, \emph{33}, 2105192\relax
\mciteBstWouldAddEndPuncttrue
\mciteSetBstMidEndSepPunct{\mcitedefaultmidpunct}
{\mcitedefaultendpunct}{\mcitedefaultseppunct}\relax
\EndOfBibitem
\bibitem[Hohlfeld \latin{et~al.}(2000)Hohlfeld, Wellershoff, G{\"{u}}dde, Conrad, J{\"{a}}hnke, and Matthias]{hohl2000}
Hohlfeld,~J.; Wellershoff,~S.-S.; G{\"{u}}dde,~J.; Conrad,~U.; J{\"{a}}hnke,~V.; Matthias,~E. {Electron and lattice dynamics following optical excitation of metals}. \emph{Chemical Physics} \textbf{2000}, \emph{251}, 237--258\relax
\mciteBstWouldAddEndPuncttrue
\mciteSetBstMidEndSepPunct{\mcitedefaultmidpunct}
{\mcitedefaultendpunct}{\mcitedefaultseppunct}\relax
\EndOfBibitem
\bibitem[Wang and Cahill(2012)Wang, and Cahill]{wang2012b}
Wang,~W.; Cahill,~D.~G. {Limits to Thermal Transport in Nanoscale Metal Bilayers due to Weak Electron-Phonon Coupling in Au and Cu}. \emph{Physical Review Letters} \textbf{2012}, \emph{109}, 175503\relax
\mciteBstWouldAddEndPuncttrue
\mciteSetBstMidEndSepPunct{\mcitedefaultmidpunct}
{\mcitedefaultendpunct}{\mcitedefaultseppunct}\relax
\EndOfBibitem
\bibitem[Chung \latin{et~al.}(2017)Chung, Wen, Huang, Gupta, Conibeer, Shrestha, Harada, and Kee]{chun2017}
Chung,~S.; Wen,~X.; Huang,~S.; Gupta,~N.; Conibeer,~G.; Shrestha,~S.; Harada,~T.; Kee,~T.~W. Nanosecond long excited state lifetimes observed in hafnium nitride. \emph{Solar Energy Materials and Solar Cells} \textbf{2017}, \emph{169}, 13--18\relax
\mciteBstWouldAddEndPuncttrue
\mciteSetBstMidEndSepPunct{\mcitedefaultmidpunct}
{\mcitedefaultendpunct}{\mcitedefaultseppunct}\relax
\EndOfBibitem
\bibitem[{Samad M. Edlou} \latin{et~al.}(1994){Samad M. Edlou}, {John C. Simons}, {Ghanim A. Al-Jumaily}, and {Nasrat A. Raouf}]{sama1994}
{Samad M. Edlou}; {John C. Simons}; {Ghanim A. Al-Jumaily}; {Nasrat A. Raouf} Optical and electrical properties of reactively sputtered TiN, ZrN, and HfN thin films. Optical Thin Films IV: New Developments. 1994; pp 96--106\relax
\mciteBstWouldAddEndPuncttrue
\mciteSetBstMidEndSepPunct{\mcitedefaultmidpunct}
{\mcitedefaultendpunct}{\mcitedefaultseppunct}\relax
\EndOfBibitem
\bibitem[R{\"o}ssle \latin{et~al.}(2021)R{\"o}ssle, Leitenberger, Reinhardt, Ko{\c{c}}, Pudell, Kwamen, and Bargheer]{ross2021}
R{\"o}ssle,~M.; Leitenberger,~W.; Reinhardt,~M.; Ko{\c{c}},~A.; Pudell,~J.; Kwamen,~C.; Bargheer,~M. The time-resolved hard X-ray diffraction endstation KMC-3 XPP at BESSY II. \emph{Journal of Synchrotron Radiation} \textbf{2021}, \emph{28}\relax
\mciteBstWouldAddEndPuncttrue
\mciteSetBstMidEndSepPunct{\mcitedefaultmidpunct}
{\mcitedefaultendpunct}{\mcitedefaultseppunct}\relax
\EndOfBibitem
\bibitem[Seo \latin{et~al.}(2004)Seo, Lee, Wen, Petrov, Greene, and Gall]{seo2004}
Seo,~H.-S.; Lee,~T.-Y.; Wen,~J.~G.; Petrov,~I.; Greene,~J.~E.; Gall,~D. Growth and physical properties of epitaxial HfN layers on MgO(001). \emph{Journal of Applied Physics} \textbf{2004}, \emph{96}, 878--884\relax
\mciteBstWouldAddEndPuncttrue
\mciteSetBstMidEndSepPunct{\mcitedefaultmidpunct}
{\mcitedefaultendpunct}{\mcitedefaultseppunct}\relax
\EndOfBibitem
\bibitem[Thomsen \latin{et~al.}(1986)Thomsen, Grahn, Maris, and Tauc]{thom1986}
Thomsen; Grahn; Maris; Tauc Surface generation and detection of phonons by picosecond light pulses. \emph{Physical review. B, Condensed matter} \textbf{1986}, \emph{34}, 4129--4138\relax
\mciteBstWouldAddEndPuncttrue
\mciteSetBstMidEndSepPunct{\mcitedefaultmidpunct}
{\mcitedefaultendpunct}{\mcitedefaultseppunct}\relax
\EndOfBibitem
\bibitem[Schick \latin{et~al.}(2014)Schick, Herzog, Bojahr, Leitenberger, Hertwig, Shayduk, and Bargheer]{schi2014b}
Schick,~D.; Herzog,~M.; Bojahr,~A.; Leitenberger,~W.; Hertwig,~A.; Shayduk,~R.; Bargheer,~M. {Ultrafast lattice response of photoexcited thin films studied by X-ray diffraction}. \emph{Structural Dynamics} \textbf{2014}, \emph{1}, 064501\relax
\mciteBstWouldAddEndPuncttrue
\mciteSetBstMidEndSepPunct{\mcitedefaultmidpunct}
{\mcitedefaultendpunct}{\mcitedefaultseppunct}\relax
\EndOfBibitem
\bibitem[DeMiguel-Ramos \latin{et~al.}(10/22/2018 - 10/25/2018)DeMiguel-Ramos, Clement, Mirea, Olivares, Felmetsger, Flewitt, and Iborra]{demira2018}
DeMiguel-Ramos,~M.; Clement,~M.; Mirea,~T.; Olivares,~J.; Felmetsger,~V.; Flewitt,~A.~J.; Iborra,~E. Hafnium Nitride as High Acoustic Impedance Material for Fully Insulating Acoustic Reflectors. 2018 IEEE International Ultrasonics Symposium (IUS). 10/22/2018 - 10/25/2018; pp 1--4\relax
\mciteBstWouldAddEndPuncttrue
\mciteSetBstMidEndSepPunct{\mcitedefaultmidpunct}
{\mcitedefaultendpunct}{\mcitedefaultseppunct}\relax
\EndOfBibitem
\bibitem[Schick(2021)]{schi2021}
Schick,~D. udkm1Dsim -- a Python toolbox for simulating 1D ultrafast dynamics in condensed matter. \emph{Computer Physics Communications} \textbf{2021}, \emph{266}, 108031\relax
\mciteBstWouldAddEndPuncttrue
\mciteSetBstMidEndSepPunct{\mcitedefaultmidpunct}
{\mcitedefaultendpunct}{\mcitedefaultseppunct}\relax
\EndOfBibitem
\bibitem[Zeuschner \latin{et~al.}(2019)Zeuschner, Parpiiev, Pezeril, Hillion, Dumesnil, Anane, Pudell, Willig, R{\"o}ssle, Herzog, von Reppert, and Bargheer]{zeus2019}
Zeuschner,~S.~P.; Parpiiev,~T.; Pezeril,~T.; Hillion,~A.; Dumesnil,~K.; Anane,~A.; Pudell,~J.; Willig,~L.; R{\"o}ssle,~M.; Herzog,~M.; von Reppert,~A.; Bargheer,~M. Tracking picosecond strain pulses in heterostructures that exhibit giant magnetostriction. \emph{Structural dynamics (Melville, N.Y.)} \textbf{2019}, \emph{6}, 024302\relax
\mciteBstWouldAddEndPuncttrue
\mciteSetBstMidEndSepPunct{\mcitedefaultmidpunct}
{\mcitedefaultendpunct}{\mcitedefaultseppunct}\relax
\EndOfBibitem
\bibitem[von Reppert \latin{et~al.}(2020)von Reppert, Mattern, Pudell, Zeuschner, Dumesnil, and Bargheer]{repp2020a}
von Reppert,~A.; Mattern,~M.; Pudell,~J.-E.; Zeuschner,~S.~P.; Dumesnil,~K.; Bargheer,~M. {Unconventional picosecond strain pulses resulting from the saturation of magnetic stress within a photoexcited rare earth layer}. \emph{Structural Dynamics} \textbf{2020}, \emph{024303}\relax
\mciteBstWouldAddEndPuncttrue
\mciteSetBstMidEndSepPunct{\mcitedefaultmidpunct}
{\mcitedefaultendpunct}{\mcitedefaultseppunct}\relax
\EndOfBibitem
\bibitem[Deb \latin{et~al.}(2021)Deb, Popova, Zeuschner, Hehn, Keller, Mangin, Malinowski, and Bargheer]{deb2021b}
Deb,~M.; Popova,~E.; Zeuschner,~S.~P.; Hehn,~M.; Keller,~N.; Mangin,~S.; Malinowski,~G.; Bargheer,~M. Generation of spin waves via spin-phonon interaction in a buried dielectric thin film. \emph{Physical Review B} \textbf{2021}, \emph{103}, 024411\relax
\mciteBstWouldAddEndPuncttrue
\mciteSetBstMidEndSepPunct{\mcitedefaultmidpunct}
{\mcitedefaultendpunct}{\mcitedefaultseppunct}\relax
\EndOfBibitem
\bibitem[Gaal \latin{et~al.}(2012)Gaal, Schick, Herzog, Bojahr, Shayduk, Goldshteyn, Navirian, Leitenberger, Vrejoiu, Khakhulin, Wulff, and Bargheer]{gaal2012}
Gaal,~P.; Schick,~D.; Herzog,~M.; Bojahr,~a.; Shayduk,~R.; Goldshteyn,~J.; Navirian,~H.~a.; Leitenberger,~W.; Vrejoiu,~I.; Khakhulin,~D.; Wulff,~M.; Bargheer,~M. {Time-domain sampling of x-ray pulses using an ultrafast sample response}. \emph{Applied Physics Letters} \textbf{2012}, \emph{101}, 243106\relax
\mciteBstWouldAddEndPuncttrue
\mciteSetBstMidEndSepPunct{\mcitedefaultmidpunct}
{\mcitedefaultendpunct}{\mcitedefaultseppunct}\relax
\EndOfBibitem
\bibitem[Gaal \latin{et~al.}(2014)Gaal, Schick, Herzog, Bojahr, Shayduk, Goldshteyn, Leitenberger, Vrejoiu, Khakhulin, Wulff, and Bargheer]{gaal2014}
Gaal,~P.; Schick,~D.; Herzog,~M.; Bojahr,~A.; Shayduk,~R.; Goldshteyn,~J.; Leitenberger,~W.; Vrejoiu,~I.; Khakhulin,~D.; Wulff,~M.; Bargheer,~M. {Ultrafast switching of hard X-rays.} \emph{Journal of synchrotron radiation} \textbf{2014}, \emph{21}, 380--5\relax
\mciteBstWouldAddEndPuncttrue
\mciteSetBstMidEndSepPunct{\mcitedefaultmidpunct}
{\mcitedefaultendpunct}{\mcitedefaultseppunct}\relax
\EndOfBibitem
\bibitem[Sander \latin{et~al.}(2016)Sander, Koc, Kwamen, Michaels, v.~Reppert, Pudell, Zamponi, Bargheer, Sellmann, Schwarzkopf, and Gaal]{sand2016}
Sander,~M.; Koc,~A.; Kwamen,~C.~T.; Michaels,~H.; v.~Reppert,~A.; Pudell,~J.; Zamponi,~F.; Bargheer,~M.; Sellmann,~J.; Schwarzkopf,~J.; Gaal,~P. {Characterization of an ultrafast Bragg-Switch for shortening hard x-ray pulses}. \emph{Journal of Applied Physics} \textbf{2016}, \emph{120}, 193101\relax
\mciteBstWouldAddEndPuncttrue
\mciteSetBstMidEndSepPunct{\mcitedefaultmidpunct}
{\mcitedefaultendpunct}{\mcitedefaultseppunct}\relax
\EndOfBibitem
\bibitem[Schick \latin{et~al.}(2014)Schick, Bojahr, Herzog, Shayduk, von {Korff Schmising}, and Bargheer]{schi2014}
Schick,~D.; Bojahr,~A.; Herzog,~M.; Shayduk,~R.; von {Korff Schmising},~C.; Bargheer,~M. udkm1Dsim---A simulation toolkit for 1D ultrafast dynamics in condensed matter. \emph{Computer Physics Communications} \textbf{2014}, \emph{185}, 651--660\relax
\mciteBstWouldAddEndPuncttrue
\mciteSetBstMidEndSepPunct{\mcitedefaultmidpunct}
{\mcitedefaultendpunct}{\mcitedefaultseppunct}\relax
\EndOfBibitem
\bibitem[Herzog \latin{et~al.}(2022)Herzog, von Reppert, Pudell, Henkel, Kronseder, Back, Maznev, and Bargheer]{herz2022}
Herzog,~M.; von Reppert,~A.; Pudell,~J.; Henkel,~C.; Kronseder,~M.; Back,~C.~H.; Maznev,~A.~A.; Bargheer,~M. {Phonon‐Dominated Energy Transport in Purely Metallic Heterostructures}. \emph{Advanced Functional Materials} \textbf{2022}, \emph{32}, 2206179\relax
\mciteBstWouldAddEndPuncttrue
\mciteSetBstMidEndSepPunct{\mcitedefaultmidpunct}
{\mcitedefaultendpunct}{\mcitedefaultseppunct}\relax
\EndOfBibitem
\bibitem[Saha \latin{et~al.}(2010)Saha, Acharya, Sands, and Waghmare]{saha2010}
Saha,~B.; Acharya,~J.; Sands,~T.~D.; Waghmare,~U.~V. Electronic structure, phonons, and thermal properties of ScN, ZrN, and HfN: A first-principles study. \emph{Journal of Applied Physics} \textbf{2010}, \emph{107}, 033715\relax
\mciteBstWouldAddEndPuncttrue
\mciteSetBstMidEndSepPunct{\mcitedefaultmidpunct}
{\mcitedefaultendpunct}{\mcitedefaultseppunct}\relax
\EndOfBibitem
\bibitem[Ginnings and Furukawa(1953)Ginnings, and Furukawa]{ginn1953}
Ginnings,~D.~C.; Furukawa,~G.~T. Heat Capacity Standards for the Range 14 to 1200°K. \emph{Journal of the American Chemical Society} \textbf{1953}, \emph{75}, 522--527\relax
\mciteBstWouldAddEndPuncttrue
\mciteSetBstMidEndSepPunct{\mcitedefaultmidpunct}
{\mcitedefaultendpunct}{\mcitedefaultseppunct}\relax
\EndOfBibitem
\bibitem[Li \latin{et~al.}(2020)Li, Wang, Hu, Gu, Tong, and Bao]{li2020}
Li,~S.; Wang,~A.; Hu,~Y.; Gu,~X.; Tong,~Z.; Bao,~H. Anomalous thermal transport in metallic transition-metal nitrides originated from strong electron--phonon interactions. \emph{Materials Today Physics} \textbf{2020}, \emph{15}, 100256\relax
\mciteBstWouldAddEndPuncttrue
\mciteSetBstMidEndSepPunct{\mcitedefaultmidpunct}
{\mcitedefaultendpunct}{\mcitedefaultseppunct}\relax
\EndOfBibitem
\bibitem[Dobrovinskaya \latin{et~al.}(2009)Dobrovinskaya, Lytvynov, and Pishchik]{dobr2009}
Dobrovinskaya,~E.~R.; Lytvynov,~L.~A.; Pishchik,~V. \emph{Sapphire. Material, Manufacturing, Applications}; Springer, 2009\relax
\mciteBstWouldAddEndPuncttrue
\mciteSetBstMidEndSepPunct{\mcitedefaultmidpunct}
{\mcitedefaultendpunct}{\mcitedefaultseppunct}\relax
\EndOfBibitem
\bibitem[Aigner \latin{et~al.}(1994)Aigner, Lengauer, Rafaja, and Ettmayer]{Aigner.1994}
Aigner,~K.; Lengauer,~W.; Rafaja,~D.; Ettmayer,~P. Lattice parameters and thermal expansion of Ti(CxN(1-x)), Zr(CxN(1-x)), Hf(CxN(1-x)) and TiN(1-x) from 298 to 1473 K as investigated by high-temperature X-ray diffraction. \emph{Journal of Alloys and Compounds} \textbf{1994}, \emph{215}, 121--126\relax
\mciteBstWouldAddEndPuncttrue
\mciteSetBstMidEndSepPunct{\mcitedefaultmidpunct}
{\mcitedefaultendpunct}{\mcitedefaultseppunct}\relax
\EndOfBibitem
\bibitem[Lucht \latin{et~al.}(2003)Lucht, Lerche, Wille, Shvyd'ko, R{\"u}ter, Gerdau, and Becker]{luch2003}
Lucht,~M.; Lerche,~M.; Wille,~H.-C.; Shvyd'ko,~Y.~V.; R{\"u}ter,~H.~D.; Gerdau,~E.; Becker,~P. Precise measurement of the lattice parameters of Al 2O3 in the temperature range 4.5--250 K using the M{\"o}ssbauer wavelength standard. \emph{Journal of Applied Crystallography} \textbf{2003}, \emph{36}, 1075--1081\relax
\mciteBstWouldAddEndPuncttrue
\mciteSetBstMidEndSepPunct{\mcitedefaultmidpunct}
{\mcitedefaultendpunct}{\mcitedefaultseppunct}\relax
\EndOfBibitem
\bibitem[{Wachtman Jr} \latin{et~al.}(1960){Wachtman Jr}, Tefft, {Lam Jr}, and Stinchfield]{wach1960}
{Wachtman Jr},~J.~B.; Tefft,~W.~E.; {Lam Jr},~D.~G.; Stinchfield,~R.~P. Elastic Constants of Synthetic Single-Crystal Corundum at Room Temperature. \emph{Journal of the American Ceramic Society} \textbf{1960}, \emph{43}, 334\relax
\mciteBstWouldAddEndPuncttrue
\mciteSetBstMidEndSepPunct{\mcitedefaultmidpunct}
{\mcitedefaultendpunct}{\mcitedefaultseppunct}\relax
\EndOfBibitem
\bibitem[von Reppert \latin{et~al.}(2018)von Reppert, Willig, Pudell, R{\"{o}}ssle, Leitenberger, Herzog, Ganss, Hellwig, and Bargheer]{repp2018}
von Reppert,~A.; Willig,~L.; Pudell,~J.-E.; R{\"{o}}ssle,~M.; Leitenberger,~W.; Herzog,~M.; Ganss,~F.; Hellwig,~O.; Bargheer,~M. {Ultrafast laser generated strain in granular and continuous FePt thin films}. \emph{Applied Physics Letters} \textbf{2018}, \emph{113}, 123101\relax
\mciteBstWouldAddEndPuncttrue
\mciteSetBstMidEndSepPunct{\mcitedefaultmidpunct}
{\mcitedefaultendpunct}{\mcitedefaultseppunct}\relax
\EndOfBibitem
\bibitem[Schick \latin{et~al.}(2012)Schick, Bojahr, Herzog, von {Korff Schmising}, Shayduk, Leitenberger, Gaal, and Bargheer]{schi2012}
Schick,~D.; Bojahr,~A.; Herzog,~M.; von {Korff Schmising},~C.; Shayduk,~R.; Leitenberger,~W.; Gaal,~P.; Bargheer,~M. Normalization schemes for ultrafast x-ray diffraction using a table-top laser-driven plasma source. \emph{The Review of scientific instruments} \textbf{2012}, \emph{83}, 025104\relax
\mciteBstWouldAddEndPuncttrue
\mciteSetBstMidEndSepPunct{\mcitedefaultmidpunct}
{\mcitedefaultendpunct}{\mcitedefaultseppunct}\relax
\EndOfBibitem
\end{mcitethebibliography}
\providecommand{\latin}[1]{#1}
\makeatletter
\providecommand{\doi}
  {\begingroup\let\do\@makeother\dospecials
  \catcode`\{=1 \catcode`\}=2 \doi@aux}
\providecommand{\doi@aux}[1]{\endgroup\texttt{#1}}
\makeatother
\providecommand*\mcitethebibliography{\thebibliography}
\csname @ifundefined\endcsname{endmcitethebibliography}  {\let\endmcitethebibliography\endthebibliography}{}

\end{document}